\documentclass[sigconf]{acmart}

\usepackage{subcaption}

\AtBeginDocument{%
  }

\setcopyright{acmlicensed}
\copyrightyear{2018}
\acmYear{2018}
\acmDOI{XXXXXXX.XXXXXXX}

\acmConference[Conference acronym 'XX]{Make sure to enter the correct
  conference title from your rights confirmation emai}{June 03--05,
  2018}{Woodstock, NY}
\acmISBN{978-1-4503-XXXX-X/18/06}

\begin{document}


\title[Resonance: Drawing from Memories to Imagine Positive Futures through AI-Augmented Journaling]{Resonance: Drawing from Memories to Imagine Positive Futures through AI-Augmented Journaling}





\author{Wazeer Zulfikar}
\affiliation{%
  \institution{MIT Media Lab}
  \city{Cambridge}
  \country{USA}
  }
\email{wazeer@media.mit.edu}

\author{Treyden Chiaravalloti}
\affiliation{%
  \institution{MIT Media Lab}
  \city{Cambridge}
  \country{USA}}
\email{treydenc@media.mit.edu}

\author{Jocelyn Shen}
\affiliation{%
  \institution{MIT Media Lab}
  \city{Cambridge}
  \country{USA}}
\email{joceshen@media.mit.edu}

\author{Rosalind Picard}
\affiliation{%
  \institution{MIT Media Lab}
  \city{Cambridge}
  \country{USA}}
\email{picard@media.mit.edu}

\author{Pattie Maes}
\affiliation{%
  \institution{MIT Media Lab}
  \city{Cambridge}
  \country{USA}}
\email{pattie@media.mit.edu}

\renewcommand{\shortauthors}{Trovato et al.}

\begin{abstract}

People inherently use experiences of their past while imagining their future, a capability that plays a crucial role in mental health. \textsc{Resonance} is an AI-powered journaling tool designed to augment this ability by offering AI-generated, action-oriented suggestions for future activities based on the user's own past memories. Suggestions are offered when a new memory is logged and are followed by a prompt for the user to imagine carrying out the suggestion. In a two-week randomized controlled study (N=55), we found that using \textsc{Resonance} significantly improved mental health outcomes, reducing the users' PHQ8 scores, a measure of current depression, and increasing their daily positive affect, particularly when they would likely act on the suggestion. Notably, the effectiveness of the suggestions was higher when they were personal, novel, and referenced the user's logged memories. Finally, through open-ended feedback, we discuss the factors that encouraged or hindered the use of the tool.


\end{abstract}

\ccsdesc[300]{Human-centered computing~Human computer interaction (HCI)}
\ccsdesc[300]{Applied computing}


\keywords{memory augmentation, mental health, large language models, positive psychology}

\begin{teaserfigure}
  \includegraphics[width=\textwidth]{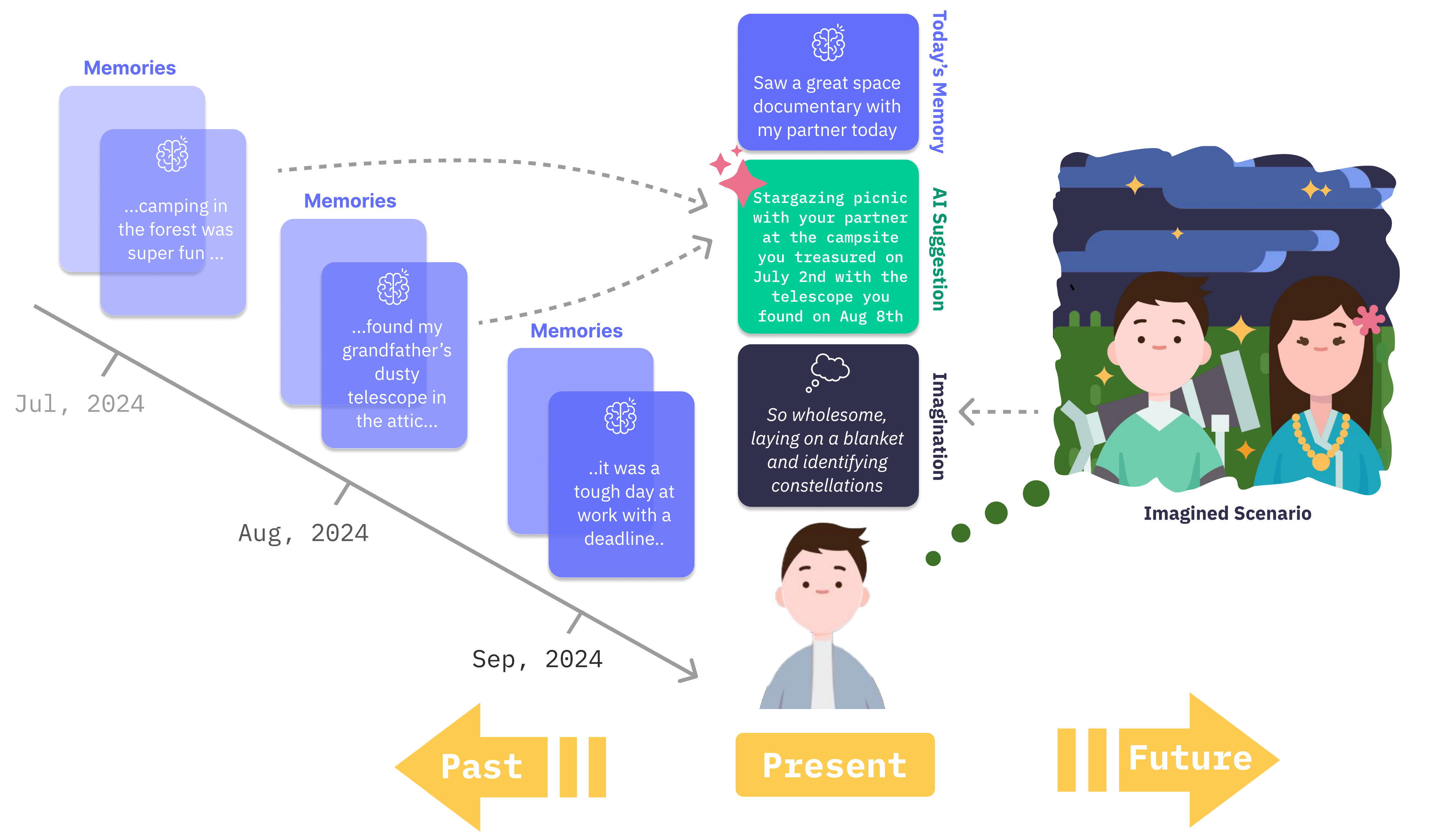}
  \caption{Flow of using Resonance to connect a user's past, present and future through memories, AI suggestions, and imaginations.}
  \Description{Flow diagram of using Resonance. User inputs a daily memory, and gets an AI suggestion (shown in the center). The AI suggestion is for a future activity, drawing from many memories (shown in the left). The user then imagines the stargazing activity (shown in the right) and journals it.}
  \label{fig:teaser}
\end{teaserfigure}

\received{20 February 2007}
\received[revised]{12 March 2009}
\received[accepted]{5 June 2009}

\maketitle

\section{Introduction}

People have the extraordinary ability to subjectively connect various aspects of their lives across different time points through mental time travel (MTT) \cite{sedikides_self_continuity_2023, tulving_episodic_2002}, the process of reflecting on the past or imagining the future. This cognitive ability is underpinned by a shared neural network between episodic memory, the conscious recollection of past episodes, and the simulation of future episodes \cite{suddendorf_mental_2009, schacter_cognitive_2007}. MTT is crucial for mental wellbeing \cite{quoidbach2009back, suo_culture_2022}, and is demonstrated by the concept of \textit{savoring} \cite{bryant_savoring_2007} in psychology. When practicing savoring, individuals attend, appreciate, and enhance positive experiences from their past, present, and future. People who regularly practice savoring report higher levels of happiness and lower levels of depression \cite{smith_savoring_2015, kurtz_savoring_2017, hurley_results_2012, jose_does_2012}.
A natural lead-in to practicing these concepts daily is through journaling, which offers a means for individuals to externalize their experiences, reminisce and emotionally process them, and plan for future scenarios \cite{koziol_journalings_2021, ullrich_journaling_2002, baikie2005emotional,sohal2022efficacy,fredrickson2010positivity, gortner2006benefits, cosley_experiences_2012}. 

With the rise in mental health struggles \cite{richter_is_2019}, growing efforts in research and commercial products look towards integrating artificial intelligence (AI) in journaling to augment our natural capacity to reflect. AI can enhance journaling by providing personalized prompts \cite{nepal2024contextual}, tailored insights based on patterns \cite{mindsera2024}, and assisting in the practice of gratitude \cite{kim2024}. However, few works have explicitly explored how AI-augmented journaling applications can strengthen the relationship between an individual's past, present, and future to enhance mental health benefits. To enhance this temporal relationship for an AI journal user, there is a need for a more holistic interface design so the AI application can effectively understand a user's needs. We define a holistic interface in this context as one that can simultaneously enable an intelligent (1)  integration with the past (i.e revisiting relevant memories to notice patterns and be grateful), and (2) preparation for the future (i.e visualizing new experiences). These aspects, when seamlessly coupled, can augment a user's need to connecting more deeply with their past and present, to believe in a more positive future, heightening the benefits of savoring experiences.

We designed and developed \textsc{Resonance}, a novel, AI-powered journal implementing a holistic interface that draws from a user's past memories to generate suggestions for future activities. When a user inputs a journal entry, the tool uses a large language model (LLM) to generate an action-oriented suggestion for future activities. The suggestions are aimed at eliciting positive emotions from the user if they are acted upon. They are personalized to the user and actively reference relevant positive moments from their memories. The user is then asked to imagine carrying out the activity suggestion, which aids in savoring the past references incorporated into the suggestion. Users were asked to journal about their imagined scenario to explain the feelings these suggestions evoked and encourage savoring for future activities. Thus, incorporating positive references from memories in the suggestion assists the user's ability to have a sense of gratitude in the present, and imagining the future suggestion facilitates creative visualization for a positive future --- practices both shown to improve mental health \cite{jans-beken_gratitude_2020, kanter_what_2010, sheldon_how_2006, gamble_futures_2021}. 

For example, as shown in Figure \ref{fig:teaser}, \textsc{Resonance} can learn that a user likes nature through a previous journal entry about their camping trip, as well as an entry about their grandfather's telescope. Then, several days later, when the user talks about having watched a space documentary with their partner, the system can suggest a stargazing picnic to increase the feeling of awe and bonding with the partner by using references from the past. Therefore, the application intuitively augments our natural capacity to make connections between the past and present to aid our imagination of positive futures.

To evaluate the \textsc{Resonance} system, we conducted a longitudinal randomized controlled study (N=55)  over 2 weeks to answer the following research questions:

\begin{itemize}
    \item \textbf{RQ1.} What effects does AI-augmented journaling with \textsc{Resonance} have on user's mental health?
    \item \textbf{RQ2.} 
    How do the features of the AI-generated activity suggestions and imagination of carrying out the suggestions, affect the estimated likeliness to act on suggestions and the user's affective state?
    \item \textbf{RQ3.} From open-ended feedback, what features may enhance the long-term use of such tools, and what barriers might prevent users from doing so? 
\end{itemize}


Participants were randomly assigned to an experimental condition where users had access to the \textsc{Resonance} tool via a web application, also accessible on a phone, or to a control condition where users could only journal on the same interface without access to the AI features of \textsc{Resonance}. All participants were asked to journal at least once a day. We collected self-report surveys throughout the study to assess mental health using the PHQ8 \cite{kroenke2009phq}, prior savoring beliefs with the SBI \cite{bryant_savoring_2003}, daily affect, and participants' open-ended feedback about the tool. The participants inputted a total of $659$ daily entries over the two weeks, with $331$ from the experimental group. Through quantitative and qualitative analyses, we find that:

\begin{itemize}
    \item PHQ8 scores, a measure of current depression, of users significantly reduced after using \textsc{Resonance} for two weeks, and the effect was independent of the users' prior beliefs in their capacity to savor.
    \item Likeliness of the user to act on the AI-generated suggestions influenced the increase in daily positive affect. The likeliness to act was correlated with the suggestion being personal, novel and containing references to the user's past. 
    \item Users appreciated the helpfulness of the AI-generated suggestions and found the imagination process to be enjoyable, however, they were concerned about privacy, and sharing and imagining difficult events.
\end{itemize}

Our findings offer insights regarding how AI-generated suggestions that draw upon memories can bridge the past and future for users. Ultimately, this work points to promising future directions in understanding how AI-augmented journaling can boost mental health by intuitively helping a user by learn from the past to envision a more positive future.

\section{Related Work}

\subsection{Mental Time Travel and Wellbeing}

Extensive research demonstrates that people can exert a large degree of control over their mental wellbeing and happiness \cite{lyubomirsky_why_2001, lyubomirsky_pursuing_2005, kaufmann_positive_2019, quoidbach_positive_2010}. One such way is through savoring \cite{bryant_savoring_2007}, coined by Bryant and Vernoff, defined as the capacity of individuals to attend, appreciate, and enhance positive experiences, and shown to drive depression levels lower and increase happiness \cite{kurtz2017savoring}. \citet{bryant_savoring_2003} further designed the Savoring Beliefs Inventory (SBI) to measure an individual's beliefs of their own capacity to savor and was validated in extensive studies \cite{lai_validation_2023, golay2018}. \citet{jose_does_2012} used a daily diary study to show that people derived happiness from positive events through savoring. Similarly, research on reminiscence and anticipation highlights their role in enhancing well-being through mental time travel \cite{bohlmeijer_effects_2007, luo_well-being_2018}. \citet{cosley_experiences_2012} found that Pensieve, a tool initiating reminiscence, not only fostered emotional connection but also empowered users to reflect on past positive experiences, amplifying their emotional resilience. \citet{isaacs_echoes_2013} explored the effect of Technology-Mediated Reflection (TMR) through a application called Echo which improved user wellbeing by enabling reflection on past digitized memories of the user. TMR was further shown to be closely linked with mood changes where when participants reflected on memories with valences opposite to their current mood, their mood became more neutral\cite{konrad_technology_2016}. \citet{macleod_wellbeing_2005} demonstrated that anticipating positive future events contributes to increased life satisfaction and a stronger sense of purpose.

Broaden-and-Build theory \cite{fredrickson2004broaden} posits that experiencing positive emotions increases the thought-action repertoire with broadened awareness. Reflecting positively on the past and maintaining a hopeful outlook on the future can build personal resources, such as resilience and social connections, which contribute to overall well-being. This can initiate and sustain an upward spiral of actions and positive emotions for individuals. \citet{gamble_futures_2021} further linked goal-directed imagination with lower depressive symptoms, where the emotional valence of goal-directed imagination strongly predicted well-being. \citet{quoidbach2009back} studied the effect of daily mental time travel, correlating purposeful engagement with mental time travel to wellbeing. Finally, \citet{schacter_cognitive_2007} discovered that there is a neurological connection between episodic memories and the imagination of future episodes, which has further been used to improve subjective well-being \cite{jing_preparing_2017}.

Recent advances in LLMs, specifically language understanding \cite{zhao_survey_2023} and idea generation \cite{girotra_ideas_2023}, enhance the potential for integrating psychological theories into daily journaling. Our work explores how AI-augmented journaling can connect a user’s past, present, and future through AI-generated suggestions grounded in the theories of savoring, reminiscence, and anticipation. \textsc{Resonance} asks users to imagine carrying out activity suggestions that would evoke positive emotions in the user. The LLM-generated suggestions actively reference positive moments from past memories to bridge different time points of a user's life. 

\subsection{AI-augmented Journaling}

The concept of using a digital medium as an external memory aid was conceived in Vannevar Bush's Memex in 1945 \cite{think1945vannevar}. Journaling, through recording thoughts, experiences, and feelings, helps people to remember experiences more profoundly \cite{donaghy1998new, finley_memory_2018}, obtain mental health benefits through reduced anxiety and stress \cite{pennebaker_forming_1999, flinchbaugh2012student, sohal2022efficacy, mercer2010visual}, improve critical and reflective thinking \cite{mccarty2020integrating, hiemstra2001uses, farrell2013teacher} and enhance learning in educational contexts \cite{raterink2016reflective, moon2006learning, stanton_using_2017}. 

Current LLMs enable systems to generate human-like utterances \cite{zhao_survey_2023} and learn in-context about user needs and preferences \cite{lampinen_can_2022, zulfikar_memoro_2024}. These new capabilities have given rise to various works incorporating AI into journaling, most of which involve scaffolding the reflection process delivered through different modalities, such as applications, videos, and chat interfaces \cite{angenius2022talkus}.  For example, \citet{kim2024} designed a system that assists users in journal writing by generating sentences to encourage diverse reflections. \citet{torres_design_2024} showed the use of AI generated ``future-self'' videos in improving emotional regulation. Similarly \citet{pataranutaporn_future_2024} showed that users reported decreased anxiety and increased future self-continuity after interacting with an AI-powered virtual version of their future selves. \citet{kim2024mindfuldiary} developed MindfulDiary to promote depth of journal entries written by psychiatric patients, and integrated expert collaboration with caregivers. \citet{angenius_design_2023} laid out design principles, such as emphasizing the user's need to be able to express in a judgment-free manner, for interactive and reflective journaling with AI.

These works highlight the growing interest in AI-assisted journaling and its impact on reflection, memory, and wellbeing. However, existing applications lack seamless integration of past and future experiences.  Our system bridges this gap by leveraging LLMs' creative capabilities and semantic search to connect users' past entries with future visualizations, adapting over time. By examining the design of \textsc{Resonance}, we offer insights into how AI-generated suggestions based on past journals can enhance mental health and inform future AI-augmented journaling systems.

\section{System Design}

\begin{figure*}
    \centering
  \includegraphics[width=0.7\linewidth]{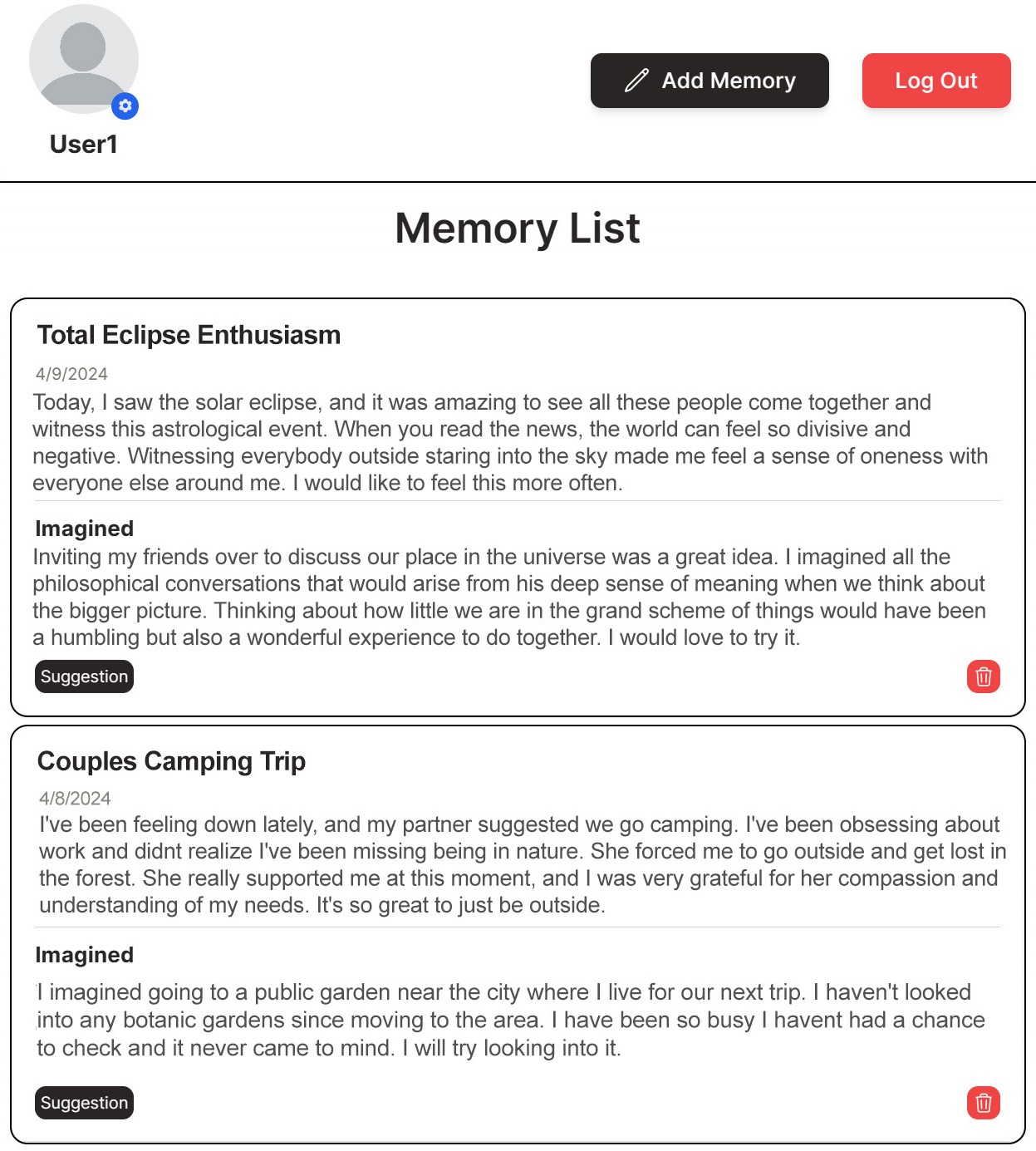}
  \caption{Dashboard of \textsc{Resonance} displaying memories and corresponding imaginations}
  \label{fig:dashboard}
\end{figure*}

\textsc{Resonance} is an AI-powered journal implemented as a web application where users can log in with credentials, enter memories, and interact with AI-generated activity suggestions. It consists of a dashboard displaying memories and a button to input a new memory from the last 24 hours as a journal entry. The interface to add a new memory can be seen in Appendix \ref{appendix:tool}, and the dashboard in Figure \ref{fig:dashboard}. After inputting a new memory, the user receives an AI-generated suggestion. The suggestion contains an activity, related to the inputted memory, that the user can do in the future. The suggested activity is aimed to evoke a positive emotion in the user if carried out. The suggestion also contains references to positive moments from memories, in the form of quotes, that are relevant to the activity, thereby personalizing the suggestion. The system avoids referencing activities in memories associated with negative emotions, which might trigger distress or discomfort. After the user has read the suggestion, they are asked to imagine carrying out the suggested activity for 30 seconds and describe the imagined scenario in the same way as a new memory. These descriptions of imagined scenarios are displayed alongside the associated memory on the dashboard. The large language model used in our system was OpenAI's GPT-4 (\textsc{gpt-4-1106-preview})\cite{openai_gpt_4_2024}. The workflow of the system is shown in Appendix \ref{appendix:tool}

\subsection{Inputting Memories}

Users can log on to the application and input a memory of an experience from the previous 24 hours. The memory is inputted using speech or text. If speech is used, the audio is transcribed to editable text and is logged as a journal entry. For a reference to that memory, a three-word, LLM-generated title is displayed along with the memory on the dashboard. The prompt used for title generation is in Appendix \ref{appendix:prompts}.

To encourage users to focus on the episodic elements while describing their memory, which include the people present, sensory perceptions, the time, and the place where the event occurred, a guideline is displayed before the user starts inputting the daily memory. The guideline is adopted from a psychology study that involves people being instructed to describe episodic memories \cite{zaman2022autonoetic}, as shown in Figure \ref{fig:resonance_new_memory}. 

\subsection{Suggestion Generator}

\begin{figure*}
    \centering
  \includegraphics[width=0.6\linewidth]{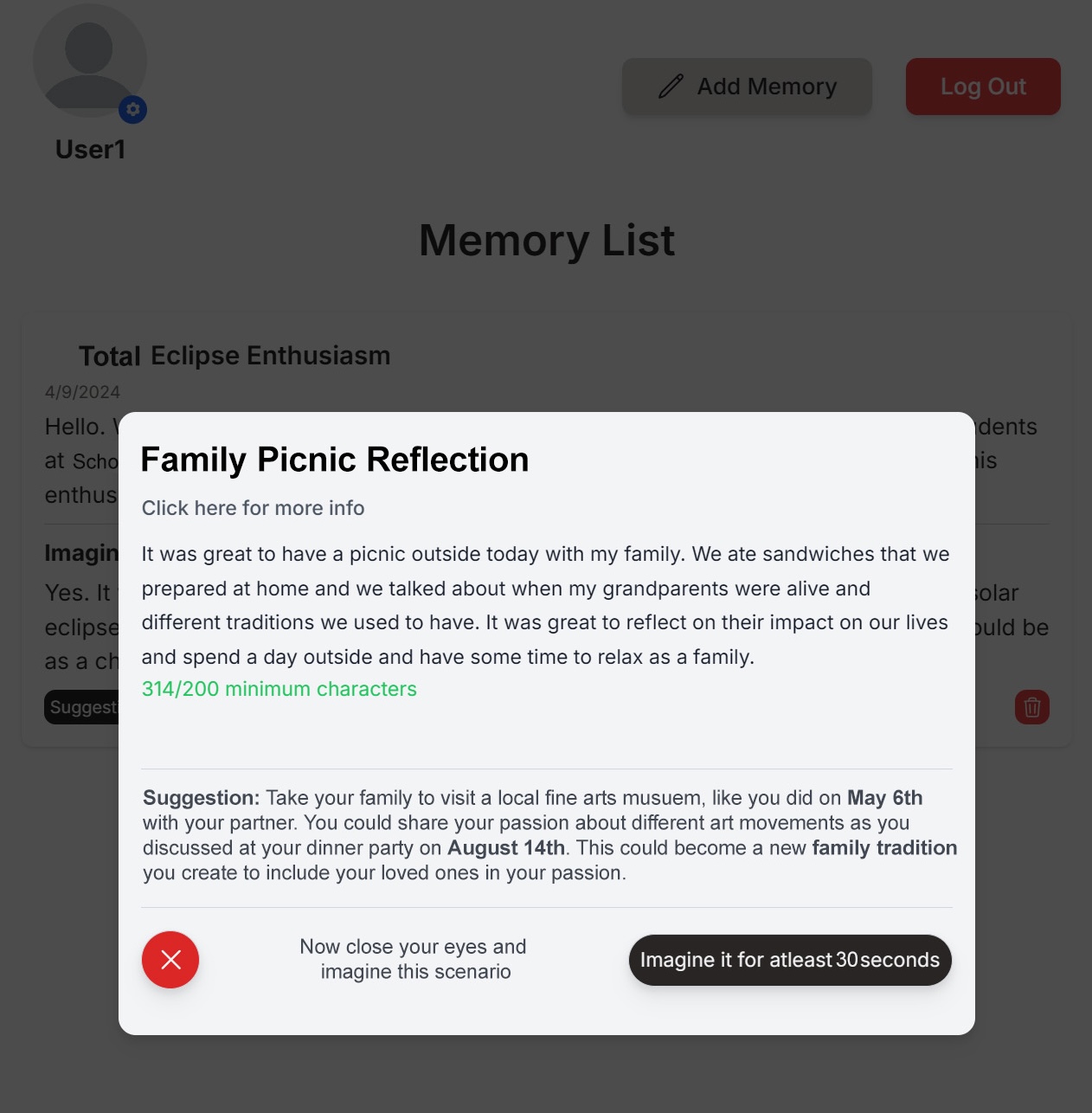}
  \caption{Example suggestion from Resonance}
  \label{fig:resonance_suggestion}
\end{figure*}

The suggestion aims to expand the user's awareness of how they can experience more positive emotions through an activity related to the logged memory for the day. LLMs were used to generate suggestions that the users might not have come up with. By referencing the user's memories, the LLM creatively bridged the user's past experiences with the present, helping them learn from the past while provoking their imagination of unexpected but relatable future activities. Delivered as a brief, two-sentence paragraph, the suggestion is designed to be personalized, novel, and positively framed. An example is shown in Figure \ref{fig:resonance_suggestion}. These core principles were crafted through prompt engineering, as detailed below. With two weeks of memories logged during the study duration, the growing context enabled suggestions to be more personalized to the specific users' needs. All the prompts used with the LLM for the following subsections are available in the Appendix \ref{appendix:prompts}. The formatting was integrated into the prompt, such as to quote the user while referencing memories. Important parts were instructed to be in bold format for emphasis. This stylistic guidance of the generated answer was crucial in making it more human-readable, deduced through the iterative testing amongst the researchers.

\subsubsection{Personalization of the suggestion}
The suggestion needs to be personalized to make it achievable for the user while reminding them of past, related moments for gratitude. To make it personal, the system retrieves semantically related memories of that particular user. The semantic similarity is calculated through the cosine similarity of the \textsc{Ada-2 embeddings} (obtained from OpenAI) of each memory with the logged memory. The five most similar memories, i.e five highest cosine similarity scores, are formatted into the prompt for generating the suggestion. The prompt is also designed to have the suggestion explicitly cite memories if they are relevant to the suggestion. Thus, the citations remind the user of positive moments from their past. This method is derived from retrieval augmented generation, which is generally used to ground LLMs with reference data \cite{lewis2020retrieval}.

\subsubsection{Novelty of the suggestion}
The novelty of suggestions to the user is instrumental in broadening their awareness of the activities that can be done in daily life to evoke more positive emotions leading to happiness. The novelty was induced by adding all the past suggestions, that the system has made for that particular user, to the prompt and designing the prompt to ensure the new suggestion to be generated had not already been suggested before. This was a variation of few-shot prompting \cite{brown_language_2020} where negative examples guided the suggestion generation.

\subsubsection{Positive framing of the suggestion}
The suggestions need to evoke positive emotions in the user. To do this effectively, prompt chaining with LLMs was used \cite{wu_promptchainer_2022}. First, the newly inputted memory was used to generate a positive emotion that could be elicited in the future while being relevant given that memory. This targeted positive emotion was then used to generate an explicit suggestion that would help evoke the specific emotion if the activity was done. Past memories referenced in the new suggestion were only chosen if relevant to the targeted positive emotion. The separate steps act as a guardrail to reduce the likelihood of suggestions that might evoke negative emotions for the user or reference a distressing memory. The study researchers checked the suggestions daily to ensure no negative suggestions, such as self-harm, were given to the participants.\citet{wu_ai_2022} showed, through a user study, how chaining significantly enhanced system transparency, controllability, and a sense of collaboration.

\subsection{Imagination of the Suggestions}
After the suggestion is offered to the user, they are asked to explicitly imagine carrying out the suggestion for 30 seconds,  intended to aid in savoring the future and past memories. The imagination process bridges the user's present, past, and future. The references to memories in the suggestion help users imagine the scenarios due to the close connection between episodic memories and future thinking \cite{schacter_cognitive_2007}. Then, the user is asked to describe the episodic elements of their imagined scenario like how their daily memory is inputted. This description allows the user to clarify how this imagination task affected them and how realistic or likely they were to act upon this suggestion. The prompt for this activity is also adopted from \citet{zaman2022autonoetic}'s study on people describing episodic memories and future thinking (as shown in Appendix \ref{appendix:tool}). The descriptions of the imagined scenarios are also saved and displayed alongside the associated memory (see \ref{fig:dashboard}).

\section{User Study}
A two-week online randomized control trial evaluated the effects of \textsc{Resonance} on real-world users. Participants were asked to use the tool daily for the two-week study duration.

\subsection{Participants}

Participants were recruited through a university behavioral research lab with a list of participants across the United States. A total of $71$ participants started the study, of which $55$ participants completed all required components of the 2-week study. Completion of the study was determined through submissions of pre and post-study surveys and at least $80\%$ of daily entries. Participants were compensated upon completion. The dropout was mainly due to the study requirement for daily journaling, i.e., inactive participants tended to drop out.  Reminders to journal were sent by email every 4 days. The demographic distribution of the $55$ participants was: 22 male, 31 female, 2 non-binary, ages = $18$ to $57$ (mean=27.4$, $ SD = 6.5).  Participants rated their prior journaling frequency between `Never' (11), `Rarely' (20), `Once a week' (7), `Several times a week' (7) and `Daily' (10). Participants also rated how useful they perceive journaling between `Not at all' (2), `Slightly' (8), `Moderately' (19), `Very' (13), and `Extremely' (13). The study was reviewed and approved by our university's ethics board. We note that most participants in the pre-study had mild depression on average, and we did not have any mental health screening criteria (minimum PHQ-8). We did not include these screening criteria, as our system was designed for the general population and not specifically for people suffering from clinical depression, which might require additional considerations described in future work.

\subsection{Conditions}
We conducted a between-subjects study, where participants were exposed to one of two conditions for the two-week period. As journaling itself can boost a person's wellbeing \citet{koziol_journalings_2021}, we designed a control condition that acted as the frame of comparison. Participants in the control condition used the same interface and had the same onboarding procedure. However, unlike our experimental condition, participants in the control did not receive AI suggestions and subsequent prompting of imagination of the suggestions. Furthermore, to minimize other confounds related to the interface, participants in the control condition had access to the same supplemental features as the experimental condition, such as reviewing their previous journal entries etc. Given that \textsc{Resonance}'s goal is to intersect the imagination process and AI suggestions, we designed our study to primarily manipulate the presence of the AI suggestions and subsequent imagination vs a journaling-only condition. While other possible controls exist, such as journaling with pre-defined suggestions, suggestions that are not contextualized to the user's journal entry could introduce variance (for example, always asking the user to imagine exercising after any journal entry). We account for the lack of this control condition by disentangling the effects of imagination and AI suggestions via user surveys and our qualitative analysis. 

\subsection{Procedure}

\subsubsection{Start of Study Surveys and Onboarding}

After obtaining consent, a set of surveys was administered through an online platform to collect demographic data and assess the baseline mental state of each participant. The baseline included measurement of current depression through the standard PHQ8 survey \cite{kroenke2009phq}. Given that the PHQ8 is designed to assess a participant's mental state over the preceding two weeks, this measure was particularly well-suited to the study’s timeline. As participants can differ in their prior beliefs on their capacity to savor, a Savoring Beliefs Inventory (SBI) questionnaire \cite{bryant_savoring_2003} was administered to evaluate the interaction it has in the mental health outcomes. During onboarding, participants were prompted for five seed memories covering pre-defined topics, such as a travel experience and a family tradition. The seed memories served as a foundation for the personalized AI suggestion generator during the initial study days. The seed memories also acted as non-recent memories for the participant, contrasting the daily memories. The questions for the seed memories are in Appendix \ref{appendix:seed_memory}.

\subsubsection{Daily Memories and Surveys}

On each day of the study, participants were required to describe a meaningful event that had occurred in the past 24 hours. A minimum of one memory was requested, but users were allowed to input more than one memory. The interface to add a memory is shown in Figure \ref{fig:resonance_new_memory}. A minimum of 200 English characters was required for each memory. To measure the daily affect of participants, each time a user started inputting a memory into the application, a two-question survey was administered to gauge the levels of how positive and negative they were feeling at that moment. We collected them using a 5-point Likert scale to the question ``How positive/negative are you feeling right now?''. Immediately after recording the memory (and interaction with the AI in the experimental condition), participants completed the same two-question survey. In the experimental conditions, participants additionally filled out a single 5-point Likert scale of how likely they would act on that generated suggestion. The exact survey questions we used are in Appendix \ref{appendix:surveys}.

\subsubsection{End of Study Surveys}

After two weeks of using \textsc{Resonance}, each participant was asked to complete an online survey to measure the state of their current mental health. The survey was administered 24 hours after the completion of the study. It was the same as the baseline measurement at the start of the study, which was the PHQ8. Additionally, participants in the experimental condition were asked to fill out surveys to assess their experiences with the AI suggestions and the imagination process. The exact questions are available in Appendix \ref{appendix:surveys}. Finally, participants from both conditions were asked to give open-ended feedback on their overall experience with the tool.

\section{Results}

We detail the analysis of data collected from the user study on the changes in mental health outcomes and user perceptions and feedback on the AI features in this section. A discussion of the results with respect to the research questions is in the following section. Participants were divided into two conditions with $N=28$ in the experimental condition and $N=27$ in the control condition. A total of $275$ onboarding memories and $659$ ($M=12.2$, $SD=2.3$) journaled daily memories were collected from the $55$ participants, with $331$ ($M=12.3$, $SD=2.5$) of these journal memories from participants in the experimental condition. Text analysis of the logged memories and imaginations are in the Appendix \ref{appendix:text_analysis}.

\subsection{Measures of Mental Health}

There were two measures of mental health used (i) the PHQ8 survey \cite{kroenke2009phq}, a standard measure of current depression, to gauge the mental health of the participants at the beginning and end of the study period, and (ii) the daily affect to gauge the affect of the participant before and after using the tool daily. We report how usage of \textsc{Resonance} impacted user's mental health outcomes (RQ1), and how this was moderated by their prior beliefs on their capacity to savor.

\subsubsection{PHQ8 Score}

The PHQ8 scores for the \textsc{Resonance} group were $M=6.11$, $SD=4.35$ before the study, and $M=4.96$, $SD=3.92$ after the study. In contrast, for the control group, they were $M=6.11$, $SD=5.09$ before the study and $M=6.14$, $SD=5.25$ after the study. A mixed effects model was used to account for repeated measures within subjects and individual variability, as each participant had PHQ8 scores before and after the study period. 

The mixed-effects model revealed that the change in PHQ8 scores was significant for the \textsc{Resonance} group. While the control group showed no significant change in PHQ8 scores over time $(p>0.05)$, \textbf{the \textsc{Resonance} group experienced a significant decrease in PHQ8 scores after the intervention} $(p<0.05)$, demonstrating the positive impact of using the AI features of the tool. The SBI scores of the participants in the \textsc{Resonance} group ($M=5.28, SD=0.77$) was added as a covariate in the mixed effects model, but it did not significantly predict changes in PHQ8 scores ($p>0.05$). This indicates that \textbf{SBI scores did not influence the effect of the intervention}.

\begin{figure*}
    \centering
  \includegraphics[width=0.5\linewidth]{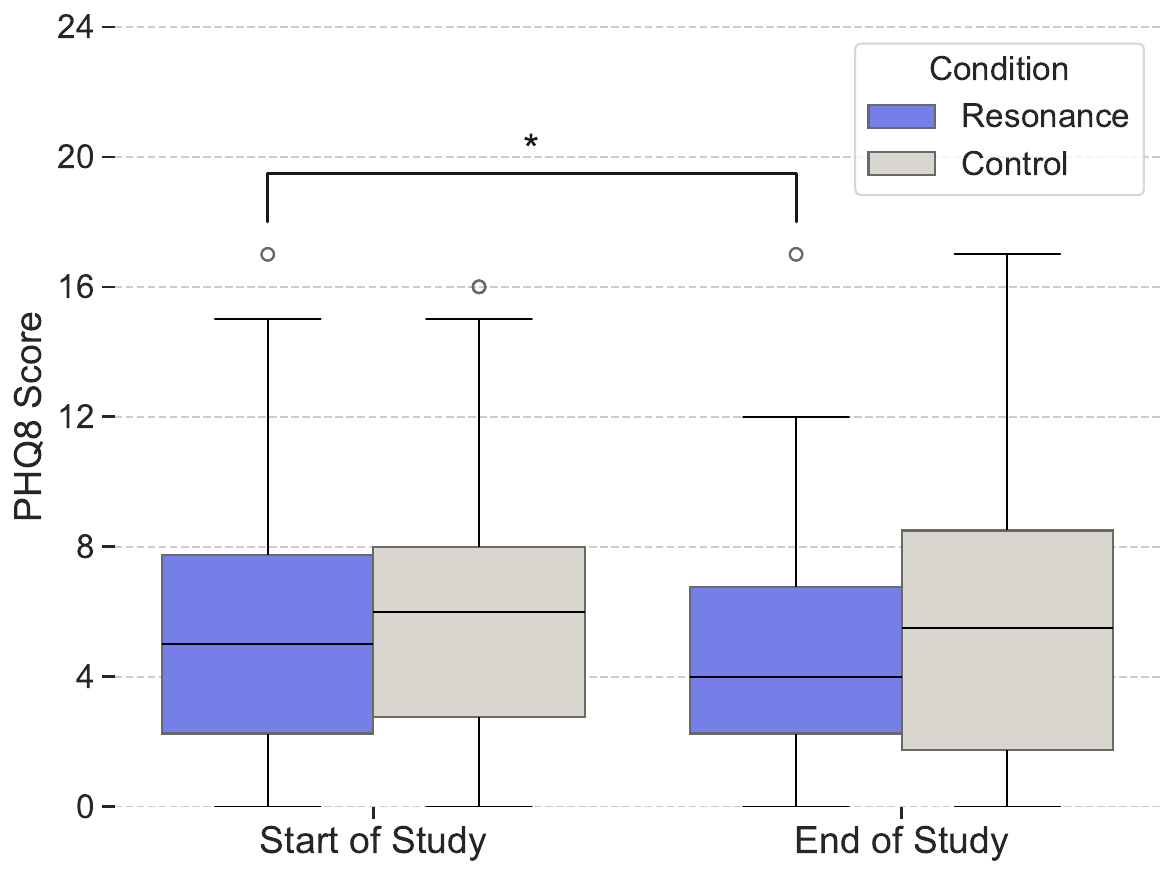}
  \caption{PHQ8 scores of participants in \textsc{Resonance} and control condition before and after the study period}
  \label{fig:phq8_change}
\end{figure*}

\subsubsection{Daily Affect Score}

\begin{figure*}[!h]
  \centering
  \subfloat[Positive Affect]{
    \includegraphics[width=0.48\textwidth]{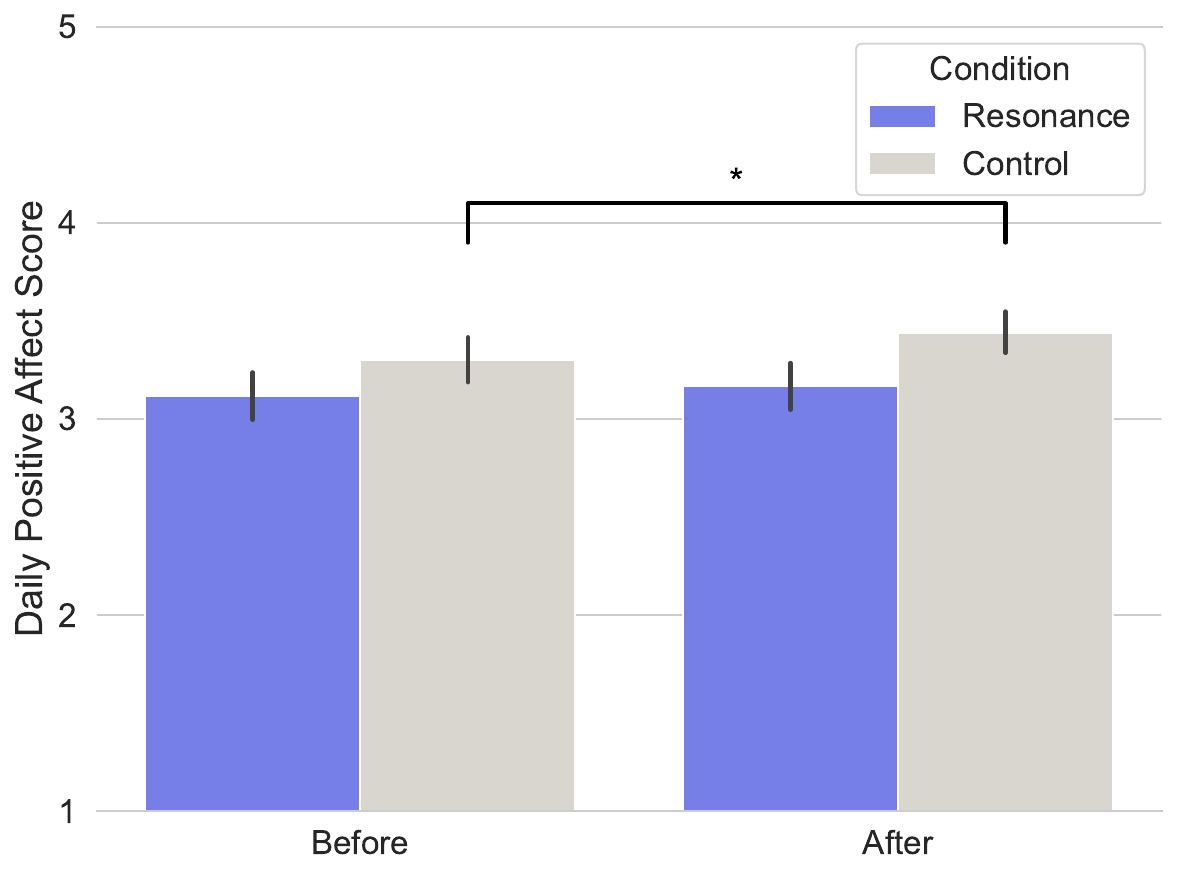}
    \label{fig:positive_affect}
  }
  \hfill
  \subfloat[Negative Affect]{
    \includegraphics[width=0.48\textwidth]{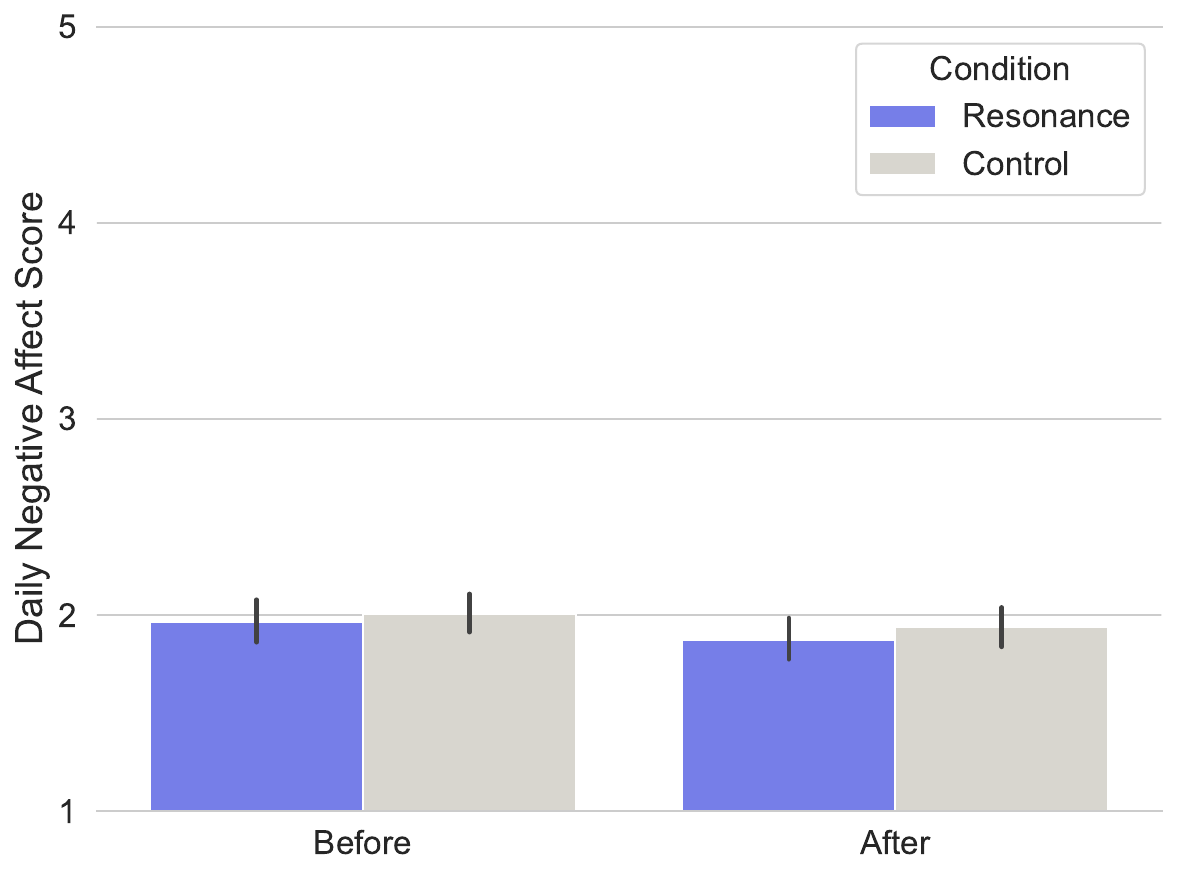}
    \label{fig:negative_affect}
  }
  \caption{Affect of participants before and after daily journaling. This included the AI interaction for the experiment condition}
  \label{fig:daily_survey_change}
  \end{figure*}

A within-subject analysis was performed for the experimental and control conditions to evaluate if there were effects on these scores, which are shown in Figure \ref{fig:daily_survey_change}. The positive affect scores for the participants in the experimental condition increased marginally from $M=3.12$, $SD=1.13$ before inputting the memory to $M=3.17$, $SD=1.10$ after inputting the imagination. The negative affect scores for the participants in the experimental condition decreased marginally from $M=1.96$, $SD=1.05$ before inputting the memory to $M=1.87$, $SD=0.98$ after inputting the memory. In comparison, the positive affect scores for the participants in the control condition increased from $M=3.30$, $SD=1.07$ to $M=3.44$, $SD=1.04$ and negative affect decreased from $M=2.01$, $SD=0.93$ to $M=1.93$, $SD=0.97$.

A mixed-effects model was used to test the statistical differences. The model revealed that there was no significant change in positive affect for the experimental condition ($p > 0.05$) but was significantly increased in the control condition ($p < 0.05$). However, there was no significant decrease in negative affect for both the experimental ($p>0.05$) and control condition ($p > 0.05$). Further analysis revealed that the positive affect before the interaction predicted the positive affect after the interaction for both the experimental ($p<0.001$) and control conditions ($p<0.001$). Similar results were found for the negative affect for both conditions ($p<.001$). These findings indicate that \textbf{the user's prior affect influences the affect of the participant after the interaction with the tool in both conditions}.

\subsection{Features of the AI Suggestions and Imaginations}
For each memory inputted by participants, one AI suggestion and a corresponding imagination was produced. We detail results related to RQ2, exploring the aspects of these two features of \textsc{Resonance}.

\subsubsection{Positivity of AI Suggestions and User Imaginations}

To measure the emotional valence of user inputs, a Linguistic Inquiry and Word Count (LIWC 2015) analysis \cite{mccarthy_applied_2012} was conducted on the logged memories, AI-generated suggestions and imagined scenarios from the participants in the experimental group to examine the use of emotional language. In the logged memories, the mean word count for positive emotion words was $M=4.02, SD=2.85$, balanced with negative emotion words at $M=1.49, SD=2.00$. In contrast, the AI-generated suggestions contained a higher mean count of positive emotion words at $M=4.34, SD=1.83$, and a lower count of negative emotion words at $M=0.19, SD=0.51$. Similarly, in participants' imagined scenarios, the word counts for positive and negative emotion words were $M=4.02, SD=2.85$ and $M=0.72, SD=1.15$, respectively. 

A mixed-effects model revealed that the positive emotion word count was significantly higher in the imagined scenarios compared to the logged memories ($p<0.05$), and the negative emotion word count was significantly lower in both the imaginations and AI-generated suggestions compared to the logged memories ($p<0.001$). Additional dimensions from the LIWC analyses describing the texts are available in the Appendix \ref{appendix:text_analysis}.

\subsubsection{Likeliness to Act on AI Suggestions}

\begin{figure*}[!h]
  \centering
  \subfloat[Positive Affect]{
    \includegraphics[width=0.48\textwidth]{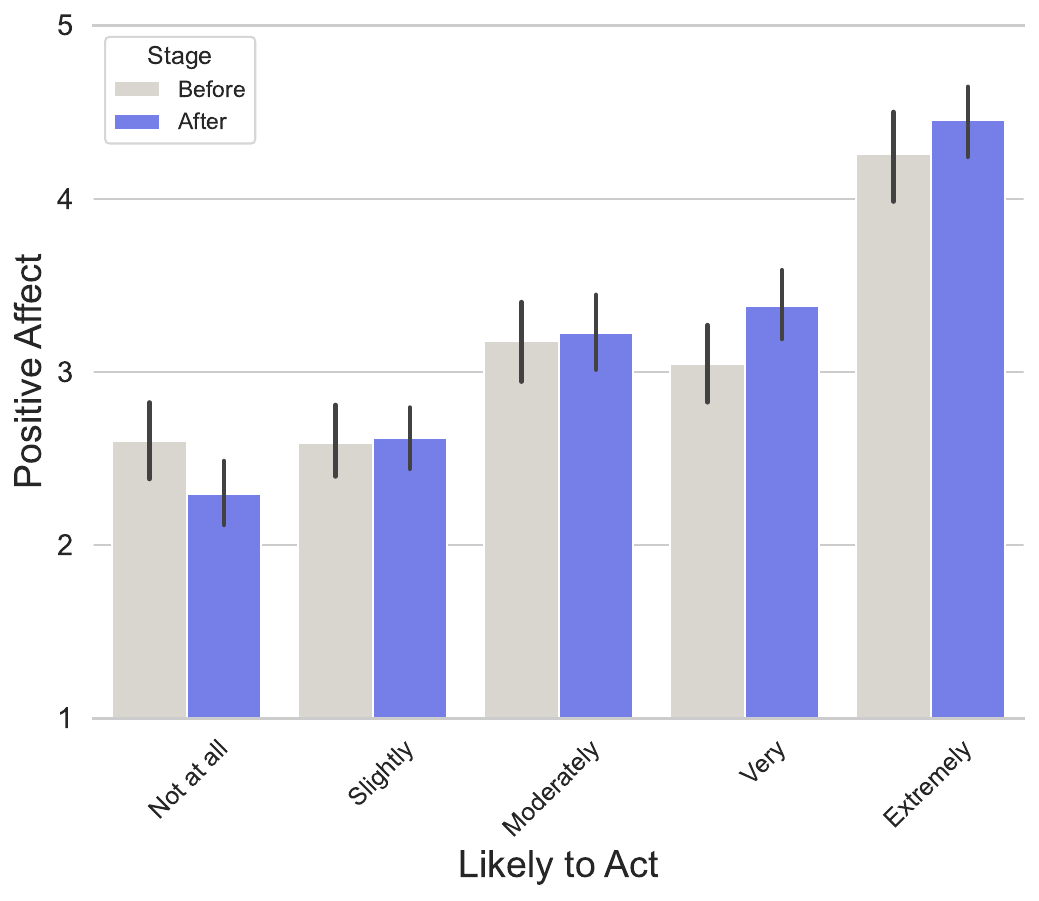}
    \label{fig:positive_affect}
  }
  \hfill
  \subfloat[Negative Affect]{
    \includegraphics[width=0.48\textwidth]{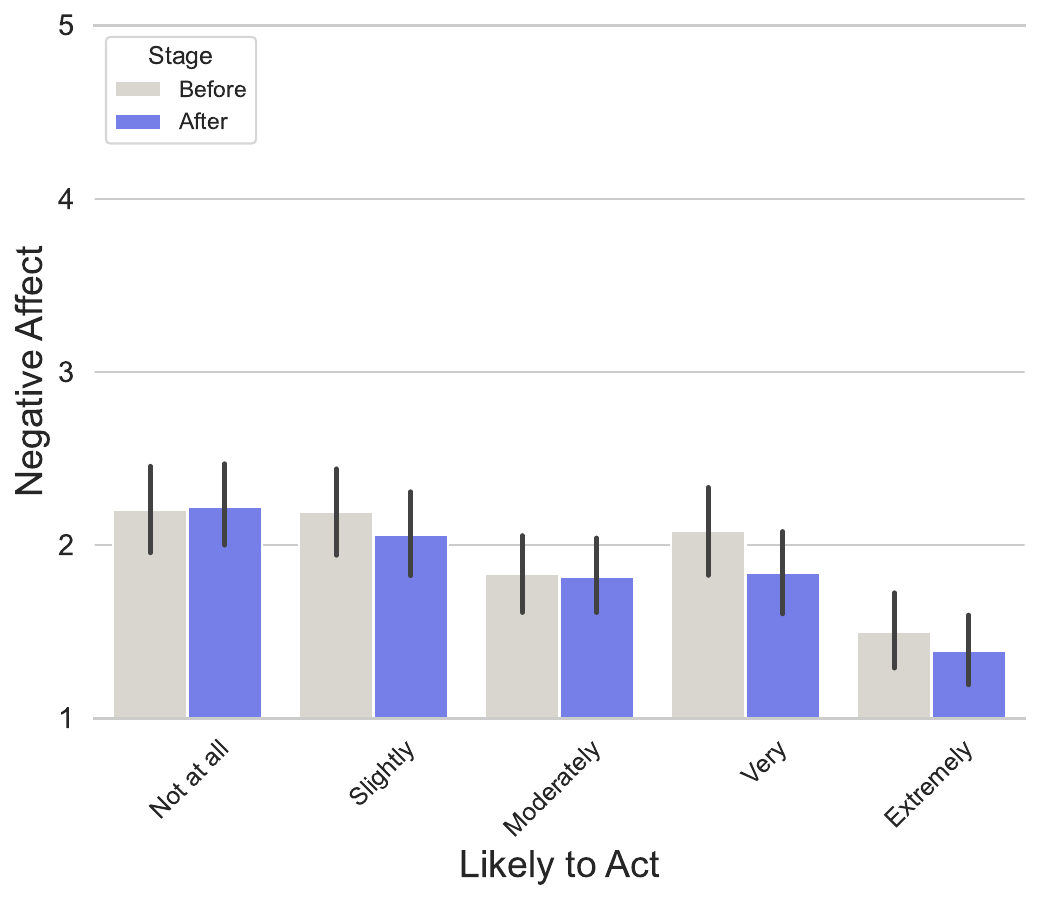}
    \label{fig:negative_affect}
  }
  \caption{Affect change of the participant before and after AI interaction conditioned on their likeliness to act on the suggestion}
  \label{fig:affect_likeliness_to_act}
  \end{figure*}

After every AI suggestion, the participants in the experimental condition rated a 5-point Likert scale of "How likely are you to act on the suggestion?" to measure the effectiveness of the suggestion to the participant. There were a total of 331 responses to the "likeliness to act on a suggestion" question, which was uniformly distributed across "Not at all" (66), "Slightly" (68), "Moderately" (72), "Very" (63) and "Extremely" (62). We analyzed how the likeliness to act on the AI suggestion impacted changes in daily affect, as shown in Figure \ref{fig:affect_likeliness_to_act}. 

The significance of the changes was tested using a mixed-effects model. The mixed-effects model was used to account for the repeated measures of affect across participants and for the individual baseline differences. The analysis revealed that \textbf{likeliness to act significantly moderated the increase in positive affect from before to after the interaction}. Participants who were more likely to act showed a greater increase in positive affect ($\beta = 0.13$, $p = 0.001$), indicating that likeliness to act was a factor in the improvement of positive affect. However, likeliness to act did not significantly affect the change in negative affect ($\beta = -0.03$, $p > 0.05$) while participants with higher likeliness to act had significantly lower negative affect overall ($\beta = -0.12$, $p = 0.001$).

\subsubsection{Perceptions of AI suggestions and Imaginations}

Twenty-four hours after the end of the 2 week study period, we measured users' perceptions of the AI-generated suggestions based on the journal memories (Appendix \ref{appendix:surveys}) and having to imagine the suggestions (Appendix \ref{appendix:surveys}). These measures were also taken on a 7-point Likert scale (1=strongly disagree, 7=strongly agree). We used a Wilcoxon signed rank test to determine if there is a significant difference (either toward agreement or disagreement with the statement) from neutral (4).  \textbf{Participants significantly agreed that the suggestion was unique to their memory and broadened their awareness of actions for increased happiness. They also disagreed that the references in the suggestion made it harder to get over difficult memories}, indicating that the AI was able to avoid drawing from difficult memories.

\subsubsection{Aspects Affecting the Effectiveness of the Suggestions}
We analyzed what aspects of suggestions made users more likely to act, as well as the role of imagination in this process, and how both these features contribute to changes in affective states of users. A Spearman correlation analysis was conducted on the mean likeliness to act, aggregated over the daily measurements for each participant, with their self-reported perceptions of the suggestions and the process of imagination. Figures \ref{fig:correlation_suggestion_likely_to_act} and \ref{fig:correlation_imagination_likely_to_act} in Appendix \ref{appendix:effectiveness} show the correlation values. \textbf{A significant correlation was found between likeliness to act on the suggestion and if the suggestion was unique to the user's memory that day, if the suggestion itself made them feel better, and if the references in the suggestions made them grateful for their past}. Similarly, a correlation was found with whether the imagination made them feel better, and if the references helped the imagination process.

\subsection{General Feedback}

At the end of the study period, participants in the experimental condition were asked the following open-ended questions to collect feedback: (i) Do you have comments about getting the AI suggestions? (ii) Do you have comments about having to imagine the AI suggestions? (iii) What did you like about the tool? (iv) What were some of the concerns?. The feedback was analyzed with the method from Braun and Clarke~\cite{braun2008} to generate the following themes:

\subsubsection{Helpful but occasionally unrealistic AI suggestions}
Participants found the suggestions to be \textit{``very helpful''} (P33), for example through \textit{``suggest[ing] healthy ways to connect with myself and others''} (P24) and noting that \textit{``sometimes suggestions themselves sometimes make me feel better''} (P41).  One participant mentioned that it was  \textit{``considerably better when responding to a positive memory as opposed to a negative memory''}(P30). However, there were times when the suggestion lacked common sense such as when it suggested a participant to \textit{``bake froyo''} (P25). Furthermore, the style and scale of the suggestion was not always appropriate to the situation at hand, such as when it was \textit{``too bossy-sounding or task-oriented''} (P39) or when it \textit{``for a small event/win, it suggested to celebrate way too much''} (P61).

\subsubsection{Imagination was beneficial but needed flexibility}
Imagining carrying out the suggestion was appreciated by participants, for example, when the \textit{``imagined scenarios would make me laugh''}(P25) and as it \textit{``may have shaped the areas in my life I want to be more aware of''} (P24). It further acted like a reinforcer for a participant as {``\textit{It strengthened the memory for the day}''} (P22). However, having options for the imagination was lacking, such as when a participant \textit{``couldn't select what you wanted to imagine''} (P12) or \textit{``if I choose to move forward with the suggestion''} (P24). A few participants would have skipped imagination as they \textit{``"immediately start rationalizing why that situation would never happen''} (P25) and when it \textit{``subjects people to think about areas of their life which may be particularly difficult''} (P24).

\subsubsection{Multimodal input amongst privacy concerns}
Participants \textit{``liked imagining while talking rather than imagining while writing''} (P30) as they found that \textit{``audio transcription made things easier''} (P43) and was \textit{``more meaningful than writing''} (P8). Such preferences came with \textit{``privacy being the largest concern''} (P48) and especially when \textit{``speak(ing) out loud about these memories and events was really difficult because I live in shared spaces and have little privacy''} (P63). A particular feature that was appreciated was the AI-generated summary titles for each memory as it was \textit{``interesting''} (P43) while being a \textit{``short but relevant title''} (P61).

\section{Discussion}
We elaborate on the results that emerged from the user study related to the research questions around the mental health effects of using \textsc{Resonance}, the contributions of the AI suggestions and user imaginations, and the open-ended feedback from the participants.

\subsection{Reduction in Depression Score and Changes in Daily Affect by Using \textsc{Resonance} [RQ1]}

As engaging in savoring has been shown to improve mental health and reduce depression, we designed \textsc{Resonance} to help people be reminded of relevant past positive experiences while imagining positive future activities through an AI-generated suggestion. While a daily session with \textsc{Resonance} lasts for less than five minutes, the consistent use of the tool every day can have compounding effects on the mental health of the user. Therefore, we evaluate the system on a daily timescale, as well as over a two-week study period. 

On a daily level, there was a significant increase in positive affect for the control and not in the experimental condition. However, further analysis revealed that participants using \textsc{Resonance} who rated themselves as more likely to act on the AI-generated suggestions for that day experienced a greater improvement in positive affect. This underscores the importance of tailoring suggestions to activities that users are inclined to act on, as it appears to be a key factor in enhancing daily positive affect. The likeliness to act on the suggestion did not influence the reduction in negative affect, indicating areas of improvement in aligning the activity suggestions to the baseline affect of the user. Further, participants with higher initial positive affect were more likely to report increased positive affect after the interaction. This cyclical relationship between affect and likeliness to act aligns with the Broaden and Build theory of positive emotions \cite{fredrickson2004broaden}. 

On a two-week level, the duration of the study period, we observe a significant reduction in the PHQ8 scores \cite{kroenke2009phq}, a measure of current depression, among users of \textsc{Resonance}, from $M=6.11, SD=4.35$ to $M=4.96, SD=3.92$. In contrast, the control group did not show a significant reduction in PHQ8 scores. Furthermore, SBI scores of users did not influence the changes in their PHQ8 scores, suggesting that \textsc{Resonance} is effective across varying levels of users' prior beliefs about their savoring capacity. These results demonstrate the effectiveness of \textsc{Resonance}'s AI features in improving mental health outcomes over the two-week period.

\subsection{Role of the AI Suggestions and the Imagination Process [RQ2]}

Overall, participants had a positive attitude towards receiving the AI suggestions and imagining themselves carrying them out. The suggestions were found to be unique to each user's memory for that day, broadened users' awareness of potential actions for enhancing their happiness, and relevant to users' through references to past memories. Importantly, referencing memories in the LLM-generated suggestions did not hinder participants' ability to cope with difficult memories. 
 We found that the relevance of the suggestion to that day's memory and its ability to broaden the user's awareness were key factors in influencing the likelihood of acting on the suggestion, emphasizing personalization and novelty. Additionally, users were more likely to act on suggestions if the reference to memories elicited gratitude, aligning with research linking gratitude to improved mental well-being \cite{watkins_gratitude_2021, emmons_gratitude_2019}.

The imagination feature was included to enhance the likelihood of acting on suggestions and to boost mental health by fostering anticipation of future positive experiences. Overall, the feedback indicated that participants neither significantly agreed nor disagreed with the value of the imagination process, suggesting that further research is needed to understand its impact. Open-ended feedback (\textit{``couldn't select what you wanted to imagine''} (P12)) showed that offering imagination as an optional feature is desired, reinforcing the need to respect the autonomy of the user. However, the imagination process in \textit{Resonance} did influence the likeliness to act on a suggestion, as there was a significant positive correlation between mean likeliness to act and whether references to memories aided users to imagine more vividly. 

These findings suggest that, while the role of the imagination feature in AI-augmented journaling warrants further investigation, incorporating references to past memories impacted how imagination affected the likelihood of acting on a suggestion, highlighting the established neurological connection between episodic memory and imagination \cite{schacter_cognitive_2007}.

\subsection{General feedback and Overall Experience [RQ3]}

Open-ended feedback was collected at the end of the study period to understand what the participants liked about the tool and their concerns about the system. Broadly, they felt that the AI suggestions were helpful, while occasionally being unrealistic. The perceived helpfulness seemed to depend on the valence of the logged memory for that day, such as it `being better for positively inclined memories than negatively inclined memories.'  The suggestions also seemed to provide `healthy ways for the participant to connect with themself and others.' The lack of commonsense knowledge in LLMs \cite{li_systematic_2021} likely caused unrealistic suggestions, such as the example cited of `baking froyo'. The suggestions reinforced the memory, consistent with studies showing journaling as memory augmentation tools \cite{donaghy1998new, finley_memory_2018}, while even being enjoyable to read for a few participants. Some noted there were difficult situations where the imagination was not feasible, and therefore, they requested more flexibility, such as optionally imagining the suggestions. This is echoed by the need for reflective flexibility in AI journaling, as outlined by Angenius and Ghajargar \cite{angenius2022interactive}, where users are given autonomy in engaging with the journaling system based on their prior emotional state. 

Participants seemed to prefer to verbalize their memories and imaginations as opposed to writing as it was "more meaningful", "easier", and "faster". The participants who desired to type their memories cited confidentiality and shared living spaces. This theme highlights the need for multimodal input to AI-augmented journaling to fit the user's context. The generated titles were also largely appreciated by participants who found them "interesting", "personalized", and "relevant". The titles served as a quick identifier to revisit and refer to memories when needed, for both the user and the AI, a feature we recommended for AI-augmented journaling. Finally, a few participants expressed their concern about the privacy of digital journaling, especially with the involvement of AI, which hindered the sharing of more personal details. This could have reduced the personalization of suggestions and thereby affected the AI interactions with the \textsc{Resonance} tool.

\subsection{Broader Implications on Mental Wellbeing}

The empirical results highlighted the difference between the effects on immediate mood and depressive symptoms. While the control group had a higher improvement in immediate mood, the treatment group had a greater reduction in depressive symptoms over two weeks. The dissonance aligns with Pennebaker's work on expressive writing  \cite{pennebaker_1997} where individuals often have short-term increases in negative emotion from writing about negative experiences but have longer-term benefits in resolving symptoms. Comments from the users such as suggestions being \textit{``too bossy-sounding or task-oriented''} (P39) or when \textit{``for a small event/win, it suggested to celebrate way too much''} (P61) show how using \textit{Resonance} may not be immediately enjoyable, but plays a role in long-term symptom reduction, an important consideration as chatbots are often designed based on the user's immediate gratification \cite{brandtzaeg_why_2017}. Additionally, increasing user autonomy with optional imagination features, as suggested by P12, \textit{``couldn't select what you wanted to imagine''}, could help users bypass unfavorable or unrealistic activity suggestions.

While our study focused on the general population, this tool could support individuals with significant depressive symptoms and other mental health conditions. For example, populations with low motivation— a common challenge in depression— may require tailored adaptations. Behavioral Activation (BA), an evidence-based therapy for depression, emphasizes planning and executing rewarding activities to counteract avoidance and low mood \cite{veale_behavioural_2008, cuijpers_behavioral_2007}. Our findings that the likeliness to act on suggestions moderate improvements in users' affect and reduce PHQ8 symptoms align with BA’s principles, as imagining and engaging in specific, achievable actions can foster positive reinforcement. Further explorations can measure whether the user carried out the suggestion. User feedback can also tailor suggestions to increase their effectiveness as a BA intervention. However, for individuals with diagnosable conditions, additional safeguards may be necessary to address the risk of rumination, especially if prompts remind them of emotionally triggering memories or disturbing scenarios. Prior research on reflective systems shows the importance of adding mechanisms to identify negative emotional responses \cite{shin_beyond_2019, kim2024mindfuldiary}. Our system does not explicitly flag distress, and future work could explore building robust safeguards, integrating sentiment analysis, and customizable prompts for avoiding distressing memories while generating suggestions to aid populations with clinical needs. Relatedly, the malleability of human memory through suggestive questioning and reconsolidation during recall needs to be considered\cite{loftus1996memory, straube_overview_2012}. An automated system such as\textsc{Resonance} could unintentionally distort logged memories and have ethical implications for the users.

\section{Limitations and Future Work}

While we show how using \textsc{Resonance} improved mental health, we discuss the following limitations in the design and study of AI-augmented journaling with Resonance.

Firstly, digital tools, such as \textsc{Resonance} that collect personal data are subject to privacy concerns because of potential data leaks. While our system uses encryption while storing memories, future work can build more privacy-enabling features, such as using a local LLM for on-device processing or de-identification of information. Secondly, the surveys used to measure the affect and the mental health of the participants could be limiting. PHQ8 was used to measure the current depression and the daily affect survey used a single Likert scale question for the positive and negative affect. More in-depth evaluation using affect measures such as  Positive and Negative Affect Schedule (PANAS), or the Self-Assessment Manikin (SAM) and human flourishing measures such as the Warwick-Edinburgh Mental Well-being Scale (WEMWBS) \cite{tennant_warwick_edinburgh_2007, watson_development_1988, bradley_measuring_1994} could be conducted in future studies. Furthermore, the act of self-reporting mental health and affect measurements could result in measurement reactivity \cite{french_reactivity_2010}, bringing events into user's awareness and subsequently impacting their mental health. 

Third, \textsc{Resonance} currently does not learn from user feedback, as it offers suggestions only based on the previously logged memories. An improvement would be to create a feedback loop where the AI learns from the user’s interaction with the tool over time. For instance, whether the user acted on a previous suggestion or not, and how much they liked imagining carrying out that suggestion can be incorporated for further personalization. Future implementations of \textsc{Resonance} on wearables with context awareness through physiological data while providing AI suggestions could be an exciting avenue to increase personalization to the user's current state. Relatedly, the positive framing of the suggestions can be ensured by having suggestions pass through an objective sentiment analysis tool where a minimum positive sentiment can be set. 

Finally, two weeks can be a limited duration to assess the long-term effects of using such a tool—especially in the context of mental health. Our study primarily serves to initially validate the effects of imagination and AI suggestions on mental wellbeing, but we point out that improvements in mental wellbeing could be related to novelty effects  \cite{laban_building_2024, shin_beyond_2019}. Moreover, the contribution of the process of imagination can be isolated in the study design to further understand its value and improve its role. It is essential to consider that imagination could have negative implications on the user, such as discouraging users to pursue actions or creating a sense of inadequacy. To address these limitations, future studies of our system will incorporate a larger population over a few months to study long-term effects with follow-up surveys after the study, with a control condition to assess the imagination process, to assess user adherence and sustained mental health improvements.

\section{Conclusion}

We designed and developed \textsc{Resonance}, a novel AI-powered journaling tool to augment the ability of people to draw upon their past experiences to envision more positive future activities. The tool prompts users to imagine carrying out AI-generated, action-oriented activity suggestions that are personalized to them. Through a randomized controlled study (N=55), we found that using the tool daily for two weeks significantly reduced PHQ8 scores, a measure of current depression, of users, regardless of their prior beliefs in savoring capacity. Further, the user's daily affect improved when they were likely to act on the AI-generated suggestions. The likeliness to act was correlated with how novel it was, and how well it referenced the user's past experiences. Finally, we found that users with privacy concerns may limit their usage but still seek the flexibility to engage with AI features, which could ultimately encourage broader adoption of AI-augmented journaling tools designed to improve mental health.

\bibliographystyle{ACM-Reference-Format}
\bibliography{sample-base}


\appendix

\section{Interface and Workflow of Tool}
\label{appendix:tool}

\begin{figure}[H]
  \centering
  \begin{subfigure}[b]{0.48\textwidth}
    \centering
    \includegraphics[width=\textwidth]{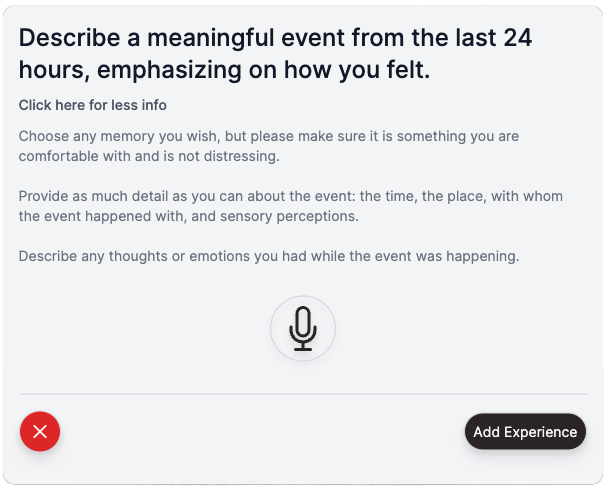}
    \caption{Adding a new memory}
    \label{fig:resonance_new_memory}
  \end{subfigure}
  \hfill
  \begin{subfigure}[b]{0.48\textwidth}
    \centering
    \includegraphics[width=\textwidth]{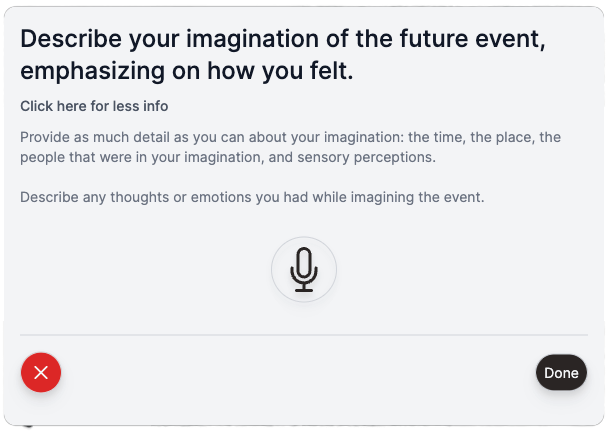}
    \caption{Adding a new imagination}
    \label{fig:resonance_new_imagination}
  \end{subfigure}
  \caption{Adding a new entry to Resonance}
  \label{fig:resonance_new_entry}
\end{figure}

\begin{figure}[H]
    \centering
  \includegraphics[width=0.9\linewidth]{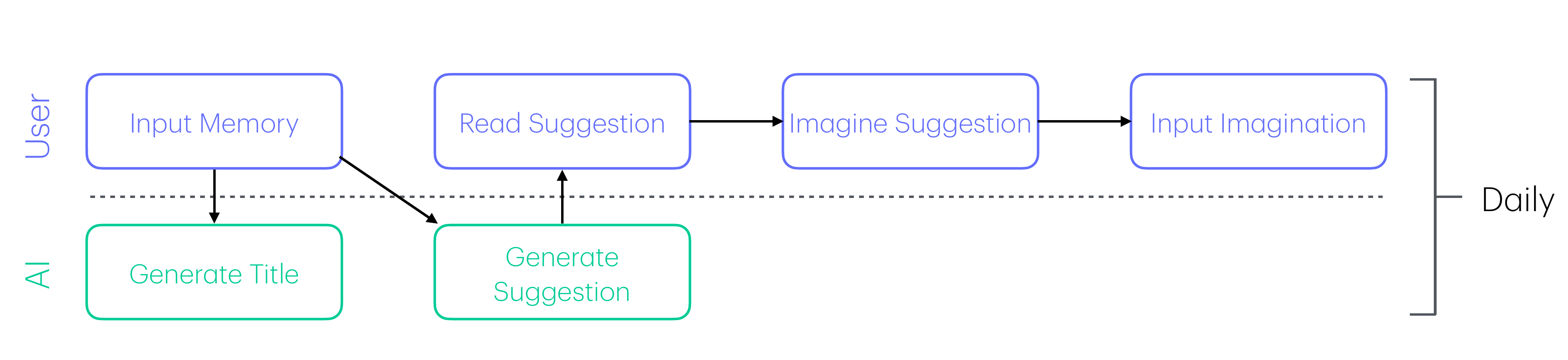}
  \caption{Workflow of \textsc{Resonance} showing the actions of the user and AI}
  \label{fig:workflow}
\end{figure}

\section{Prompts}
\label{appendix:prompts}

\subsection{LLM Prompt of Title Generation}
\begin{quote}
{\small \fontfamily{qcr}\selectfont
You are a title generator for journal entries. Create a 3 word title that accurately captures the entry and would be unique. Don't use quotes.
}
\end{quote}

\subsection{LLM Prompt for Positive Emotion}
\begin{quote}
{\small \fontfamily{qcr}\selectfont
You are a helpful assistant that helps me feel a positive emotion based on an experience of mine today.\\
Today's Experience: \{new memory\} \\
Given today's experience above, what is one suggestion for a positive emotion that could be elicited or increased in intensity? \\
Give only the emotion, describe it and why it would be good for me. Be creative. \\
Suggestion (less than 40 words):
}
\end{quote}

\subsection{LLM Prompt for Action Suggestion}
\begin{quote}
{\small \fontfamily{qcr}\selectfont
You are a helpful assistant that helps me reflect on my memories to influence my future experience more positively.\\
Related Past Memories: \{memories\} \\
Today's Experience: \{new memory\} \\
Today's date: \{current datetime\} \\
What is a easy and doable action related to today's experience that will definitely make me feel more \{emotion\}? 
Imagine the action while integrating this with elements of the past related memories.
Be creative, concise and personal. End with how it will help feel \{emotion\}. Back up the answers with references to memories if needed, by citing and quoting them. 
Dates should of the format of 23rd Nov.
Give only the suggestion. You can make important parts bold text using <b>bold</b>. \\
Suggestion (less than 60 words):
}
\end{quote}


\section{Seed Memory Questions}
\label{appendix:seed_memory}

The following were the questions for the seed memory during onboarding of the participants.

\begin{enumerate}
    \item Describe a travel experience that deeply moved or changed you. What was it about this experience that was so impactful?
    \item What has been the biggest challenge you've faced in life, and how did you overcome it? What did you learn from this experience?
    \item Is there a particular cultural event or practice you’ve experienced that left a lasting impression on you?
    \item What is your most cherished memory from your childhood, and why does it stand out to you?
    \item Can you recall a family tradition that you particularly loved? How did it shape your understanding of family?
\end{enumerate}

\section{Surveys}
\label{appendix:surveys}

\subsection{Daily Survey}

We measured two aspects using a 5-point Likert scale (1=Not at all, 2=Slightly, 3=Moderately, 4=Very, 5=Extremely). 

\begin{itemize}
    \item How positive are you feeling?
    \item How negative are you feeling?
\end{itemize}

If given an AI-generated action suggestion, then they were also asked the following question with the same 5-point Likert scale.

\begin{itemize}
    \item How likely are you to act on this suggestion?
\end{itemize}

\subsection{End of Study Survey - Perceptions}

The perceptions of the participants towards the suggestions and imaginations were gauged using 7-point Likert scales (1=strongly disagree, 7=strongly agree).

Please read carefully and answer the following statements truthfully. These statements are related to the AI suggestions and not the imagination.

\begin{itemize}
    \item The suggestions were unique to my memory each day
    \item The suggestions made me feel better about that memory
    \item The suggestions broadened my awareness of what I can do for my happiness
    \item The suggestions were repetitive
    \item The suggestions were irrelevant to what I can do for my happiness
    \item The references to past memories in the suggestion made the suggestion more relevant to me
    \item The references to past memories in the suggestion made me more grateful about my past
    \item The references to past memories in the suggestion made it harder to get over difficult memories
    \item The suggestions increased the types of memories I inputted in the following days
    \item I would have liked to spend time conversing with the AI further receiving the suggestion
    \item The suggestions made that current memory more meaningful
\end{itemize}

Please read carefully and answer the following statements truthfully. These statements are related to the imaginations after the AI suggestions.

\begin{itemize}
    \item Imagining the suggestions made me feel better about that memory
    \item Imagining the suggestions made it more likely that I will act on it
    \item Imagining the suggestions was a waste of my time
    \item I enjoyed imagining a suggestion even if I was not going to act on that suggestion
    \item If I wanted to act on the suggestion, then I would have preferred to not have to imagine it
    \item The references to past memories in the suggestions helped me imagine the suggestion more vividly
    \item Imagining the suggestions increased the types of memories I inputted in the following days
    \item Imagining the suggestion made me look forward to it more, if I was to act on it.
\end{itemize}

\subsection{End of Study Survey - Open Ended}

\begin{itemize}
    \item What did you like about the tool?
    \item What concerns do you have with such a tool?
    \item Is there anything else you would like to share about your experience?
\end{itemize}

\subsection{Survey Results}

\begin{table}[htbp]
\centering
\caption{Participant feedback on AI suggestions (1=strongly disagree, 7=strongly agree). * indicates significantly different from neutral  (p<0.05)}
\begin{tabular}{p{0.6\columnwidth}cc}
\toprule
\textbf{Statement} & \textbf{Mean (M)} & \textbf{Std (SD)} \\ 
\midrule
The suggestions were unique to my memory each day$^*$ & 5.57 & 1.43 \\ 
The suggestions made me feel better about that memory & 4.11 & 1.89 \\ 
The suggestions broadened my awareness of what I can do for my happiness$^*$ & 4.93 & 1.72 \\ 
The suggestions were repetitive (R) & 4.39 & 1.71 \\ 
The suggestions were irrelevant to what I can do for my happiness (R)$^*$ & 3.29 & 1.72 \\ 
The references to past memories in the suggestion made the suggestion more relevant to me$^*$ & 4.82 & 1.83 \\ 
The references to past memories in the suggestion made me more grateful about my past & 4.46 & 1.79 \\ 
The references to past memories in the suggestion made it harder to get over difficult memories (R)$^*$ & 2.54 & 1.57 \\ 
The suggestions increased the types of memories I inputted in the following days & 3.93 & 1.82 \\ 
I would have liked to spend time conversing with the AI further receiving the suggestion (R) & 4.11 & 1.83 \\ 
The suggestions made that current memory more meaningful$^*$ & 4.78 & 1.44 \\ 
\bottomrule
\end{tabular}

\label{tab:suggestion_feature}
\end{table}

\begin{table}[htbp]
\centering
\caption{Participant feedback on imagination of AI suggestions (1=strongly disagree, 7=strongly agree).}
\begin{tabular}{p{0.6\columnwidth}cc}
\toprule
\textbf{Statement} & \textbf{Mean (M)} & \textbf{Std (SD)} \\
\midrule

Imagining the suggestions made me feel better about that memory & 4.46 & 1.86 \\ 

Imagining the suggestions made it more likely that I will act on it & 4.29 & 2.05 \\ 
Imagining the suggestions was a waste of my time (R) & 3.57 & 1.93 \\ 
I enjoyed imagining a suggestion even if I was not going to act on that suggestion & 4.32 & 1.81 \\ 
If I wanted to act on the suggestion, then I would have preferred to not have to imagine it (R) & 3.43 & 1.85 \\ 
The references to past memories in the suggestions helped me imagine the suggestion more vividly & 4.43 & 1.95 \\ 
Imagining the suggestions increased the types of memories I inputted in the following days & 3.82 & 1.87 \\ 
Imagining the suggestion made me look forward to it more, if I was to act on it & 4.18 & 1.85 \\ 
\bottomrule
\end{tabular}

\label{tab:imagination_feature}
\end{table}

\section{Logged Memories and Imaginations}
\label{appendix:text_analysis}

\begin{table}[htbp]
\centering
\caption{Character length of memories, AI suggestions and Imaginations}
\begin{tabular}{cccc}
\toprule
\textbf{Condition} & \textbf{Type} & \textbf{Mean (M)} & \textbf{Std (SD)} \\ 
\midrule
Resonance & Memories & 490 & 295 \\ 
Resonance & AI Suggestion & 308 & 36 \\ 
Resonance & Imagination & 403 & 186 \\ 
Control & Memories & 581 & 450 \\ 
\bottomrule
\end{tabular}
\end{table}

A Linguistic Inquiry and Word Count (LIWC 2015) analysis \cite{mccarthy_applied_2012} was conducted to describe the memories, suggestions and imaginations.

\begin{table}[htbp]
\centering
\caption{LIWC word count (Affect, Social, Cognitive Processes) of text analyzed for the memories, AI suggestions and imaginations}
\begin{tabular}{llccc}
\toprule
\textbf{Condition} & \textbf{Type} & \textbf{Affect} & \textbf{Social} & \textbf{Cognitive} \\ 
\midrule
Resonance & Memories       & $5.6 \pm 3.5$  & $7.6 \pm 8.8$   & $12.1 \pm 10.2$ \\
Resonance & AI Suggestion  & $4.7 \pm 1.9$  & $6.5 \pm 2.8$   & $3.7 \pm 1.9$   \\
Resonance & Imagination    & $5.3 \pm 3.1$  & $5.0 \pm 4.2$   & $13.1 \pm 7.6$  \\
Control   & Memories       & $6.6 \pm 5.3$  & $7.4 \pm 9.4$   & $13.5 \pm 11.7$ \\
\bottomrule
\end{tabular}
\label{tab:liwc_part1}
\end{table}

\begin{table}[htbp]
\centering
\caption{LIWC word count (Perceptual, Biological Processes) of text analyzed for the memories, AI suggestions and imaginations}
\begin{tabular}{llcc}
\toprule
\textbf{Condition} & \textbf{Type} & \textbf{Perceptual} & \textbf{Biological}\\ 
\midrule
Resonance & Memories       & $3.2 \pm 2.8$  & $2.0 \pm 2.9$  \\
Resonance & AI Suggestion  & $1.4 \pm 1.2$  & $1.1 \pm 1.4$  \\
Resonance & Imagination    & $2.6 \pm 2.1$  & $1.3 \pm 1.9$  \\
Control   & Memories       & $3.6 \pm 3.4$  & $2.1 \pm 3.5$  \\
\bottomrule
\end{tabular}
\label{tab:liwc_part2}
\end{table}

\section{Aspects affecting effectiveness of suggestions}
\label{appendix:effectiveness}

\begin{figure*}
    \centering

  \includegraphics[width=\linewidth]{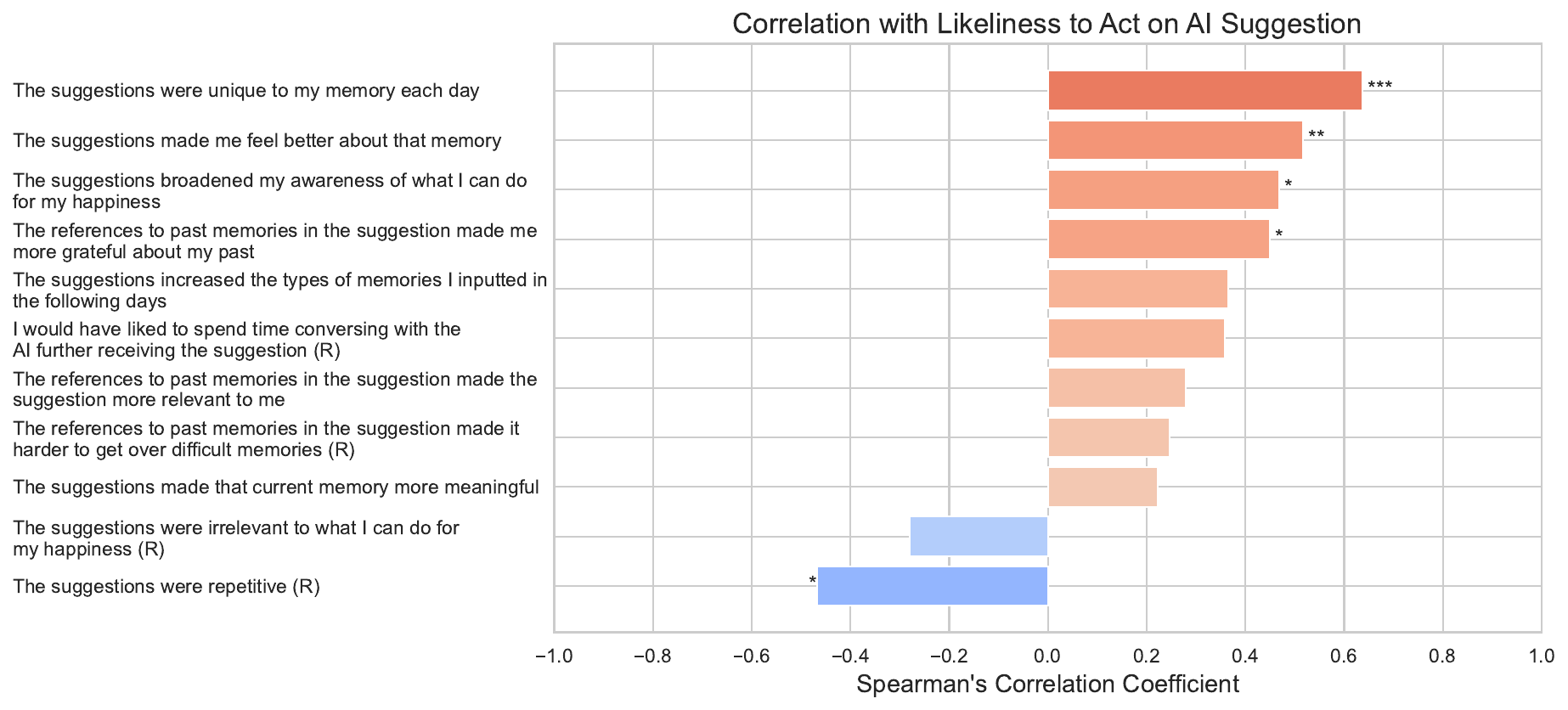}
  \caption{Correlations between mean likeliness to act of each participant and self-report perceptions of receiving the AI-generated activity suggestions}
  \label{fig:correlation_suggestion_likely_to_act}
\end{figure*}

\begin{figure*}
    \centering

  \includegraphics[width=\linewidth]{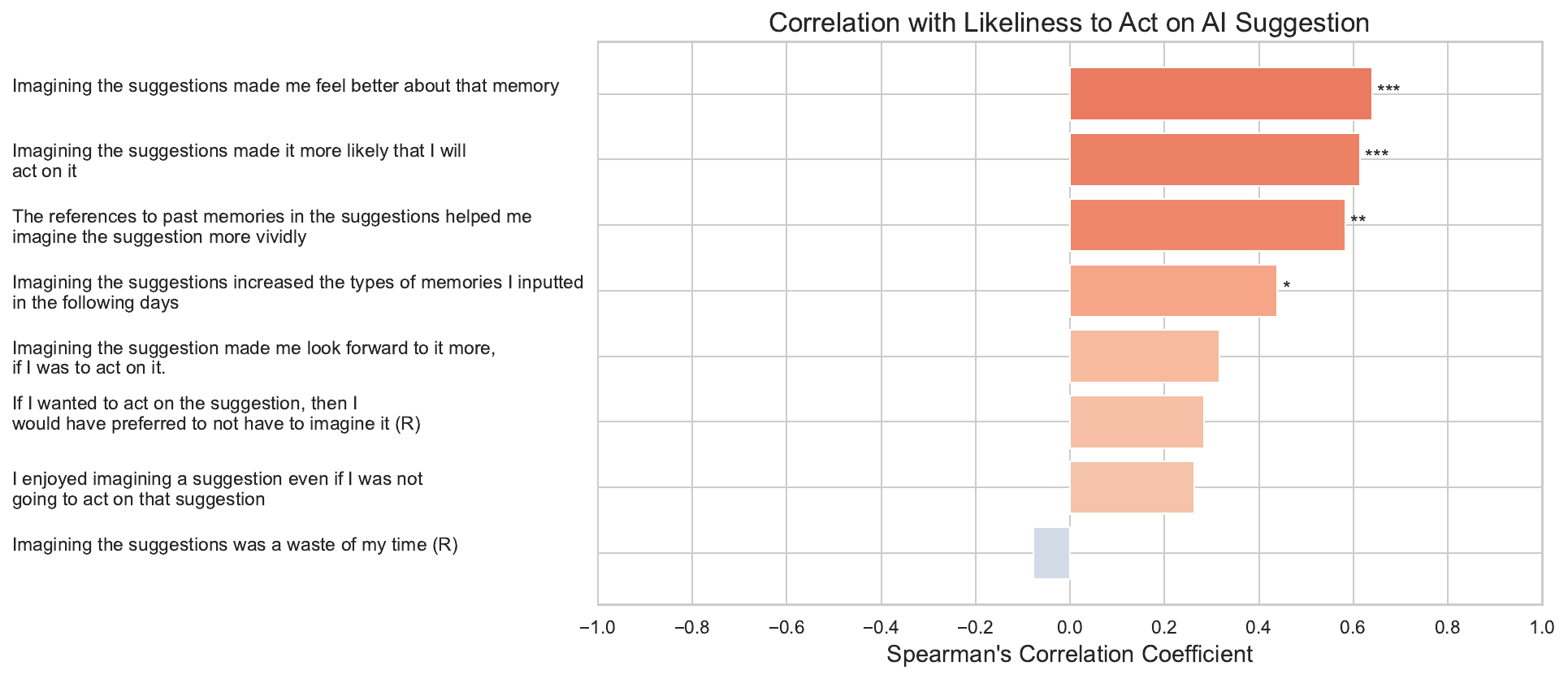}
  \caption{Correlations between mean likeliness to act of each participant and self-report perceptions of imagining the AI-generated activity suggestions}
  \label{fig:correlation_imagination_likely_to_act}
\end{figure*}

\end{document}